\documentclass{aa}
\usepackage{amsmath,graphicx,amssymb}
\usepackage{txfonts}
\usepackage{natbib}
\bibpunct{(}{)}{;}{a}{}{,}

\newlength\staretab

\makeatletter
\def\sgn{\mathop{\operator@font sgn}\nolimits}
\makeatother

\defcitealias{bohlender88}{B88}

\begin{document}

\title{Time-dependent modeling of extended thin decretion disks of critically 
rotating stars}

\author{P.~Kurf\"urst\inst{1} \and A.~Feldmeier\inst{2} \and J.~Krti\v{c}ka\inst{1}}

\institute{Department of Theoretical Physics and Astrophysics,
           Masaryk University, Kotl\'a\v rsk\' a 2, CZ-611\,37 Brno, Czech Republic
           \and
           Institut f\"ur Physik und Astronomie, Universit\"at Potsdam, Karl-Liebknecht-Stra\ss e 24/25, 
           14476 Potsdam-Golm, Germany}

\date{Received}

\abstract {During their evolution massive stars can reach the phase of critical rotation when 
a further increase in rotational speed is no longer possible.
Direct centrifugal ejection from a critically or near-critically rotating
surface forms a gaseous equatorial decretion disk.
Anomalous viscosity provides the efficient mechanism for transporting the angular momentum outwards.
The outer part of the disk can extend up to a very large distance from the parent star.} 
{We study the evolution of density, radial and azimuthal velocity, and angular momentum 
loss rate of equatorial decretion disks out to very distant regions. 
We investigate how the physical characteristics of the disk depend on
the distribution of temperature and viscosity.}
{We calculated stationary models using the Newton-Raphson method. 
For time-dependent hydrodynamic modeling we developed the numerical code based on 
an explicit finite difference scheme on an Eulerian grid including 
full Navier-Stokes shear viscosity.}
{The sonic point distance and the maximum angular momentum loss rate strongly depend on the 
temperature profile and are almost independent of viscosity.
The rotational velocity at large radii rapidly drops accordingly to 
temperature and viscosity distribution.
The total amount of disk mass and the disk angular momentum 
increase with decreasing temperature and viscosity.}
{The time-dependent one-dimensional models basically confirm the results obtained in the 
stationary models as well as the assumptions of the analytical approximations. 
Including full Navier-Stokes viscosity we systematically avoid the 
rotational velocity sign change at large radii.
The unphysical drop of the rotational velocity and angular momentum loss at
large radii (present in some models) can be avoided in the models with decreasing temperature and
viscosity.}


\titlerunning{Modelling of extended thin decretion disks of critically 
rotating stars}

\authorrunning{P.~Kurf\"urst et al.}
\maketitle

\section{Introduction}

The outflowing disks may be formed around various types of stars, such
as Be stars, B[e] stars, and possibly luminous blue 
variables \citep[e.g.,][]{Smith}
and asymptotic giant branch stars \citep[see, e.g.,][]{Matt,Ruyter}.
Observational evidence supports the idea that these disks are Keplerian (rotationally supported) 
gaseous disks \citep{Carcio}. 
The solutions for time-independent viscous decretion disk structures where one assumes the disk to be 
isothermal and in vertical hydrostatic equilibrium \citep[e.g.,][]{okazaki,Carcio}
show that, because of highly
supersonic rotational velocity, the disks are geometrically
very thin, and the disk opening angle is only a few degrees in the region
close to the star.
These studies naturally support the idea of the viscous decretion disk model leading to the formation of 
near-Keplerian disks around critically rotating stars.
However, since the specific angular momentum in the 
near-Keplerian disk increases with the radius, 
decretion disks are unlikely near-Keplerian far from 
the star. 
A natural expectation is that a disk which is Keplerian near a star
becomes angular-momentum conserving far from the star, although
this transitional feature of decretion disks
is not very well understood theoretically nor has it been observationally confirmed.
Most of the models analyze 
the inner parts of the disk, whereas the evolution close to the sonic point or even in 
the supersonic regions up to the possible outer disk edge has not been, to our knowledge,
very well studied.

In this paper we study the characteristics of the outflowing disks of
critically rotating stars.
The mass loss rate is determined by the angular momentum loss rate needed to keep the star at critical rotation 
\citep[see also \citealt{Kurf1}, \citealt{Kurf2}]{Krticka}.
The basic scenario follows the model of the viscous decretion disk proposed by 
\citet[see also \citealt{okazaki}]{Lee}. In this model \citet{Lee} obtained 
a steady structure of viscous decretion disks around Be stars in thermal and radiative equilibrium. 
As the physics of accretion disks is quite similar, we follow the main principles \citep{Pringle,Frank} 
used in accretion disk theory for the description of decretion disks.
The main uncertainties are the viscous coupling and the temperature distribution. Although some recent
models indicate a constant value of the viscosity throughout 
such disks \citep{Penna}, 
we investigate the cases when the viscous coupling varies outward as a certain power law. The disk temperature 
is mainly affected
by the irradiation from the central star \citep{Lee}. 
Motivated by the non-local thermodynamic equilibrium (NLTE) simulations \citep{Carcio} 
we extrapolate the temperature distribution up to quite distant regions as 
a power law.

\section{Basic equations and parameterization}
\label{baseq}

The disk is described in a cylindrical coordinate system assuming axial symmetry 
\citep[e.g.,][]{Lee,okazaki,Krticka}.
The mass conservation (continuity) equation in the
geometrically thin case is (we denote the cylindrical radius as $R$, the
spherical radius as $r$)
\begin{equation}
\label{massconserve}
R\frac{\partial\Sigma}{\partial t}+\frac{\partial}{\partial R}\left(R\Sigma V_{\scriptscriptstyle{R}}\right)=0,
\end{equation}
where $\Sigma$ is the
surface density defined as 
\begin{equation}
\label{surfden}
\Sigma=\int_{-\infty}^{\infty}\rho\,\text{d}z,
\end{equation}
$\rho$ is the density, and $V_{\scriptscriptstyle{R}}$ is the radial component 
of velocity. 
The equation of radial momentum conservation in the thin disk approximation is
\begin{equation}
\label{radmomconserve}
\frac{\partial V_R}{\partial t}+V_R\frac{\partial V_R}{\partial R}=
\frac{V_{\phi}^2}{R}-\frac{GM}{R^2}-\frac{1}{\Sigma}\,\frac{\partial (a^2\Sigma)}{\partial R}+\frac{3}{2}\,\frac{a^2}{R},
\end{equation}
where $V_{\scriptscriptstyle{\phi}}$ is the azimuthal component 
of velocity, $a$ is the speed of sound,
$a^2=kT/(\mu m_u)$,
$\mu$ is the mean molecular weight ($\mu=0.62$ for the ionized hydrogen gas),
$m_u$ is the atomic mass unit, and $M$ is stellar mass. 
The last term on the right-hand side comes from the description of the gravitational force in a cylindrical coordinate system  
in the thin disk approximation, i.e., the disk thickness is negligible with respect to radial distance of any point studied 
\citep{Matsu, Krticka, Kurf2}.
From the conservation of azimuthal component of momentum we have 
\begin{equation}
\label{phimcon}
\frac{\partial V_\phi}{\partial t}+V_R\frac{\partial V_\phi}{\partial R}+
\frac{V_{R}V_{\phi}}{R}=f_{\text{visc}},
\end{equation}
where $f_{\text{visc}}$ means the viscous force per unit volume,
exerted by the outer disk segment on the inner disk segment. In the axisymmetric
($\partial/\partial\phi=0$)
and geometrically thin case
the viscous force density is
$f_{\text{visc}}=\partial T_{R\phi}/\partial R+2T_{R\phi}/R$ \citep[e.g.,][]{Mihalas2}, 
where $T_{R\phi}$ denotes the $R$-$\phi$ component
of the stress tensor. Viscous force density $f_{\text{visc}}$ is
usually represented by using the first order linear viscosity term (see Eq.~\eqref{viscon})
with the adopted Shakura-Sunyaev $\alpha$~parameter \citep{Shakura}. 
We also examine the cases with non-constant $\alpha$ parameters.
Taking into account the turbulent motion of the gas, 
we can write for the kinematic viscosity $\nu$ \citep{Frank}
\begin{equation}
\label{shak}
\nu=\alpha\frac{a^2R}{V_{\phi}}\approx\alpha aH,
\end{equation}
with $H$ denoting the typical vertical scaleheight of the disk, $H^2=a^2R^3/(GM)$ 
in Keplerian case.

The NLTE simulations \citep[e.g.,][]{Carcio} show that the radial temperature distribution 
in the very inner regions (up to few stellar radii)
corresponds to a flat blackbody reprocessing disk due to the optically thick nature of this inner part,
$(T_0\approx\frac{1}{2}T_{\mathrm{eff}},\,T(r)\propto R^{-0.75})$,
where $T_0$ is the disk temperature at $R=R_{\mathrm{eq}}$. 
As the disk becomes vertically optically thin, the disk temperature rises to the
optically thin radiative equilibrium temperature with the average of about $60\%$ of $T_{\mathrm{eff}}$. 
The temperature radial profile at larger radii is nearly isothermal with
relatively mild temperature decrease (about $1000~\text{K}$ from $10$ to $50$ stellar radii).
\citet{Millar1, Millar2} also found the radial temperature distribution up to $100$ stellar radii to be nearly isothermal.
We approximate these dependencies
by a radial power law
\begin{equation}\label{temperature}
T=T_0(R_{\mathrm{eq}}/R)^p,
\end{equation}
where $p$ is a free parameter ($0\leq p<0.5$). 
Using the same power law temperature decline we also extrapolate the radial temperature structure of the 
outer part of the disk.

The viscosity also influences the temperature, but the
contribution of the viscous heating in the disk is very
small (practically negligible) compared
with the heating that comes from
radiative flux from the star \citep[e.g.,][]{smak, Carcio}. 
The viscous heating dominates
in the inner disk regions \citep{Lee} only in the
case of enormous value of mass loss rate ($\dot{M}\geq 10^{-5}\,M_{\odot}\text{yr}^{-1}$).
This is the reason 
why the viscosity is parameterized via temperature independent $\alpha$ parameter.

The radial profile of $\alpha$ is not quite certain. The common agreement is that the value of
the $\alpha$ parameter of turbulent viscosity is less than $1$, 
since for $\alpha>1$ the rapid thermalization due to shocks
would lead again to $\alpha\le 1$ \citep{Shakura}. Most authors use the value around $0.1$; some of the most recent
works focused on the disk viscosity problem \citep[e.g.,][]{Penna}
find the $\alpha$ viscosity coefficient as constant, $\alpha\approx 0.025$. 
However, here we regard it as an
open question if the $\alpha$
should be considered as constant throughout the disk or not.
Therefore, we introduce 
\begin{equation}
\label{alpvis}
\alpha=\alpha_0(R_{\mathrm{eq}}/R)^n,
\end{equation}
where $\alpha_0$ is the viscosity of the inner region of the disk near the stellar surface, 
$n$ is a free parameter of the radial viscosity dependence, $n>0$.
The kinematic viscosity $\nu(R)$ is related to 
temperature via Eq.~\eqref{shak}.

The first order linear viscosity term in Eq.~\eqref{phimcon} gives
\begin{equation}
\label{viscon}
f_{\text{visc}}^{(1)}=-\frac{1}{R^2\Sigma}\,\frac{\partial}{\partial R}(\alpha a^2R^2\Sigma),
\end{equation}
while including the full viscosity term we obtain the right-hand side of
Eq.~\eqref{phimcon} in the form 
\begin{equation}\label{angfl} 
f_{\text{visc}}^{(2)}=\frac{1}{R^2\Sigma}\,\frac{\partial}{\partial R}
\left(\alpha a^2R^3\Sigma\,
\frac{\partial\,\text{ln}\,V_{\phi}}{\partial R}-
\alpha a^2R^2\Sigma\right).
\end{equation}
The same expression is obtained by employing the angular momentum equation \citep{Pringle, Frank}
\begin{equation}\label{angfl1} 
\frac{\partial}{\partial t}\left(\Sigma R^2\Omega\right)+\frac{1}{R}\frac{\partial}{\partial R}\left(\Sigma V_RR^3\Omega\right)
=\frac{1}{2\pi R}\frac{\partial G}{\partial R} 
\end{equation}
(used in numerical scheme; see Sect.~\ref{tdepcalc}), 
where $\Omega$ is the angular velocity ($\Omega=V_{\phi}/R$) and $G$ is the viscous torque acting between two neighboring disk segments,
\begin{equation}\label{vist}
G=2\pi\alpha a^2\Sigma R^3\frac{\partial\,\text{ln}\,\Omega}{\partial R}.
\end{equation}

Most authors use the concept of some outer disk radius 
$R_\text{out}$, where, because of the 
radiation pressure for example, the disk matter may be completely driven outward \citep{Lee}.
Using the reasonable parameterization of $T$ and $\alpha$, we are not 
dependent on the choice of $R_{\text{out}}$ and can calculate the radial profiles of the characteristics as a direct 
solution of the hydrodynamic equations.

\section{Analytical estimates of the radial thin disk structure}
\label{radthin}

To obtain a general idea about the behavior of the main characteristics of the system 
we consider the stationary form of the basic hydrodynamic equations \citep{okazaki,Krticka}.
Integrating Eq.~\eqref{phimcon} (using the term $f_{\text{visc}}^{(2)}$ 
from Eq.~\eqref{angfl}) and dividing this by the 
stationary mass conservation equation ($R\Sigma V_R=\text{const.}$) 
we derive for $\alpha=\text{const.}$
\begin{equation}\label{angestim2} 
RV_{\phi}+\frac{\alpha a^2R}{V_R}\left(1-R\frac{\partial\,\text{ln}\,V_{\phi}}{\partial R}\right)=\text{const.}
\end{equation}
Inclusion of only the first order linear viscosity term (see Eq. \eqref{viscon}) 
gives a similar relation without the second term in the bracket.
In the innermost part of the disk, $V_R\ll a$. Consequently, the second term on
the left-hand side of Eq.~\eqref{angestim2} dominates, therefore $V_R\sim R$ and
$\Sigma\sim R^{-2}$ from the continuity equation \citep{okazaki}.
In the inner region of the disk the radial pressure gradient is negligible
compared with the gravitational force,
hence from the momentum equation \eqref{radmomconserve} Keplerian rotation follows,
$V_\phi\sim R^{-1/2}$. 
Since $\dot{M}=\text{const.}$ (see Eq.~\eqref{massconserve}), we have in this region
$\dot{J}(\dot{M})\sim RV_{\phi}\sim R^{1/2}$.
Eq.~\eqref{angestim2} becomes 
\begin{equation}\label{angestim3} 
RV_{\phi}+\gamma\frac{\alpha a^2R}{V_R}=\text{const.},
\end{equation}
where for Keplerian rotational velocity the numerical factor $\gamma=3/2$.
Obviously, the second order linear viscosity term in the Keplerian case represents 
one half of the corresponding first order linear term 
assuming the same $\alpha$ parameter, and therefore cannot be neglected.
In the distant region near the sonic point ($V_R\approx a$) the radial velocity increases;
therefore, the first term on the left-hand side of Eq.~\eqref{angestim2} fully
dominates, hence $RV_{\phi}=\text{const.}$
and the disk is angular momentum conserving. 

In very distant supersonic regions with nearly flat disk temperature distributions 
($V_R\gg a$, $V_R\gg V_{\phi}$, $\partial a^2/\partial R\approx 0$)
and negligible gravity the radial momentum equation \eqref{radmomconserve} with
use of mass conservation equation \eqref{massconserve} implies logarithmic
radial dependence, $V^2_R\sim \text{ln}\,R$. 
The second term on the left-hand side of Eq.~\eqref{angestim3} rises,
consequently the azimuthal velocity has to substantially decrease and may 
in general become even negative. 
This is, however, not possible according to the logarithmic term in Eq.~\eqref{angestim2}.
The numerical simulations (see Figs.~\ref{B0pt0}-\ref{ptI}) prove that the  
azimuthal velocity even in extremely distant regions does not change its direction.
Equation \eqref{angestim3} indicates that for steeper viscosity and temperature decrease (lower $\alpha$ and $a^2$ in
distant regions) the region of azimuthal velocity drop moves outwards.

From the stationary form of the radial momentum and mass conservation equations (\ref{radmomconserve}) 
and \eqref{massconserve} with the help of Eq.~(\ref{temperature}),
the sonic point condition \citep*{okazaki,Krticka} is fullfilled at the sonic point radius
\begin{equation}\label{spcon}
R_{\text{s}}=GM\left[\left(\frac{5}{2}+p\right)a^2+V_{\phi}^2\right]^{-1}.
\end{equation}
Substituting $V_{\phi}(R_\text{s})\approx\frac{1}{2}V_K(R_\text{s})$, where
$V_{K}(R)=\sqrt{GM/R}$ is the Keplerian velocity \citep[see also
Figs.~\ref{fig1} and \ref{fig2}]{Krticka} we derive an estimate of the sonic point
radius
\begin{equation}\label{sonicestR}
\frac{R_\text{s}}{R_{\text{eq}}}\approx
\left[\frac{3}{10+4p}\left(\frac{V_{K}(R_{\text{eq}})}{a(R_{\text{eq}})}\right)^2\right]^{\frac{1}{1-p}}.       
\end{equation}
Radius of the maximum of the disk angular momentum loss
roughly corresponds to the sonic point radius 
(see Figs.~\ref{fig1}-\ref{ptI}).
Since $\dot{J}_{\text{max}}\approx\dot{M}R_\text{s}V_{\phi}(R_\text{s})$, we have
\begin{equation}\label{sonicestL}
\dot{J}_{\text{max}}(\dot{M})\approx
\frac{1}{2}\left[\frac{3}{10+4p}\left(\frac{V_{K}(R_{\text{eq}})}{a(R_{\text{eq}})}\right)^2\right]^{\frac{1}{2-2p}}
\dot{M}R_{\text{eq}}V_{K}(R_{\text{eq}}).       
\end{equation}
From the above equations we can see, that adding cooling (higher $p$) can substantially increase the sonic point distance and
consequently the angular momentum loss of the disk for a fixed $\dot M$.
For example, for $p=0.2$ the maximum loss rate of angular momentum
increases roughly 2 times and for $p=0.4$ it increases roughly 5-6 times in comparison with the isothermal case, 
according to the type of star.

\section{Numerical methods}

\subsection{Stationary calculation}
\label{statcalc}

For the stationary models \citep{Kurf2} we solved the system of
Eqs.~\eqref{massconserve}, \eqref{radmomconserve}, and \eqref{phimcon} omitting the explicitly 
time-dependent terms, using the Newton-Raphson method 
\citep[e.g.,][]{krticka2003}.
For the purpose of this study we selected the star with the following parameters corresponding 
to the main-sequence spectral Type B0 \citep{Harmanec}:
${T_{\text{eff}}=30\,000\,\text{K},M=}{14.5\,M_{\odot},\,R_{\star}=5.8\,R_{\odot}}.$
To solve the system of hydrodynamic equations (supplemented by the sonic point condition Eq.~\eqref{spcon}) numerically
we used the so-called shooting method, based on changing the inner boundary (photospheric) 
radial velocity in order to find a proper branch of the solution.
The azimuthal velocity at the inner disk boundary 
(stellar equatorial surface) corresponds to Keplerian velocity. The solution of the system of hydrodynamic equations is independent of
the scaling of the surface density $\Sigma$ \citep[cf.][]{Krticka}
and the mass loss rate $\dot{M}$ is
treated as a free parameter in our calculations.
In our models the inner boundary radius is the equatorial radius of the critically rotating star,
$R_{\text{eq}}=3/2\,R_{\star}$.

Numerical problems occured when we attempted to involve the second order viscosity term according to 
Eq.~\eqref{angfl}.
Despite its complete analytic linearization in the Jacobi matrix, the solutions suffered from severe vibrations or 
perturbations mainly in the proximity of and above the sonic point, and so  
we used only the first order viscosity term Eq.~\eqref{viscon} in the azimuthal
momentum equation \eqref{phimcon}.
For the numerical calculation we selected a radial grid consisting of 300-1000 grid points according to various initial conditions.
We used the numerical package LAPACK \citep{anders} to solve
the system of linearized equations.

\subsection{Time-dependent calculation}
\label{tdepcalc}

For the time-dependent calculations we write the left-hand sides of hydrodynamic equations 
in conservative form \citep[see, e.g.,][]{Normi, Hirsch, Stony, felda, Lev1}
\begin{equation}
\frac{\partial\boldsymbol{u}}{\partial t}+\boldsymbol{\nabla}\cdot\boldsymbol{F}(\boldsymbol{u})=0, 
\end{equation}
where the quantities $\boldsymbol{u}$ $=$ $\Sigma$, $\Sigma\boldsymbol{V}$, $\boldsymbol{R}\times\Sigma\boldsymbol{V}$ 
and $\boldsymbol{F}(\boldsymbol{u})$ $=$ $\Sigma\boldsymbol{V}$, $\Sigma\boldsymbol{V}\otimes{\boldsymbol{V}}$, 
$\boldsymbol{R}\times\Sigma\boldsymbol{V}\otimes{\boldsymbol{V}}$
for the mass, momentum, and angular momentum conservation equations, respectively (with $\times$ denoting the vector product and 
$\otimes$ denoting the tensor product). 
Assuming axial symmetry of the thin disk, $\partial/\partial\phi=0$, 
all functions are only radially and time dependent.
The radial component 
of mass conservation equation is given in Eq.~\eqref{massconserve} and
the radial component of momentum equation (see Eq.~\eqref{radmomconserve}) in its conservative form gives 
\begin{equation}\label{genmom}
\frac{\partial(\Sigma V_R)}{\partial t}+\frac{1}{R}\frac{\partial}{\partial R}(R\Sigma V_R^2)
-\Sigma\frac{V_{\phi}^2}{R}=-\frac{\partial P}{\partial R}-\Sigma\frac{GM}{R^2}+\frac{3}{2}\,\frac{a^2}{R},
\end{equation}
where $P$ is the isothermal gas pressure, $P=a^2\Sigma$.
The explicit form of the angular momentum equation (see Eq.~\eqref{angfl1}) in this case is
\begin{equation}\label{genang}
\frac{\partial}{\partial t}(R\Sigma V_{\phi})+\frac{1}{R}\frac{\partial}{\partial R}(R^2\Sigma V_RV_{\phi})=
R\Sigma f_{\text{visc}}^{(2)},
\end{equation}
the term $f_{\text{visc}}^{(2)}$
denotes the density of the viscous force
in a form derived in Eq.~\eqref{angfl}. 
Because we parameterize the disk temperature profile
via Eq.~\eqref{temperature},
we do not employ the energy equation for the calculation in this case.

For the time-dependent calculations we extended the
one-dimensional hydrodynamic code of \citet{felda}.
Following \citet{Normi}, the angular momentum advection flux acts as the azimuthal component of momentum flow.
We nevertheless do not use the consistent advection 
schema \citep{Norm1}
as it is described in detail in \citet{Normi}, but employ Eq.~\eqref{angfl1} in its explicit form. 
Equations \eqref{massconserve}, \eqref{radmomconserve}, and \eqref{angfl1} with use of Eq.~\eqref{vist} are discretized using
time-explicit operator-splitting and finite difference method on staggered radial grids \citep[see][]{Lev1}. 
The advection fluxes are 
calculated on the boundaries of control volumes of these grids \citep[see, e.g.,][]{Roache, Lev2} using van Leer's monotonic interpolation 
\citep{vanleer1, vanleer2}. 

In the source steps regarding the right-hand sides of Eqs.~\eqref{genmom} and \eqref{genang}
we accelerate the fluid 
by the action of external forces (gravity) and internal pressure forces on the gas momenta \citep[see][]{Normi}. 
Involving the radial artificial viscosity term $Q$ (see Eq.~\eqref{artivis})
we solve finite-difference approximations 
to the following differential equations \citep[see][]{Normi}, 
\begin{eqnarray}  
\left.\frac{\text{d}\Sigma}{\text{d}t}\right|_{\text{source}}&=&0,\\
\left.\frac{\text{d}\Pi}{\text{d}t}\right|_{\text{source}}&=&\Sigma\frac{V_{\phi}^2}{R}
-\frac{\partial(a^2\Sigma)}{\partial R}-\Sigma\frac{GM}{R^2}+\frac{3}{2}\,\frac{a^2}{R}-\frac{\partial Q}{\partial R},\\ 
\left.\frac{\text{d}J}{\text{d}t}\right|_{\text{source}}&=&R\Sigma f_{\text{visc}}^{(2)},
\end{eqnarray}
where $\Pi=\Sigma V_R$ is the radial momentum density, $J=\Sigma RV_\phi$ is the angular momentum density,
and $f_{\text{visc}}^{(2)}$ is the second order viscosity term derived in Eq.~\eqref{angfl}.
We adopt the artificial viscosity $Q$ in the explicit form \citep[see also, e.g., \citealt{Caramana}]{Normi} 
\begin{equation}\label{artivis}
Q_i=\Sigma_i(V_{R,\,i+1}-V_{R,\,i})[-\text{C}_1a+\text{C}_2\text{min}(V_{R,\,i+1}-V_{R,\,i},0)],
\end{equation}
where $V_R$ is the radial velocity component, $a$ is the sound speed, the
lower index $i$ denotes the $i$-th spatial grid step.
The second term scaled by a constant $\text{C}_2=1.0$ is the quadratic artificial viscosity \citep[see][]{Caramana} used in compressive zones. 
The linear viscosity term should be sparingly used for damping 
oscillations in stagnant regions of the flow \citep{Normi}. We use this term with $\text{C}_1=0.5$ rarely when some oscillations may occur
near the inner boundary region (near stellar equatorial surface), either in cases with low $\alpha_0$ viscosity parameter 
($\alpha_0<0.02$) or in cases with steeper temperature decrease ($p>0.2$).

For the time-dependent modelling we employ two types of stars: main sequence stars of spectral Type B0 \citep{Harmanec} 
with parameters introduced in Sect.~\ref{statmod} and a Pop III star with the following parameters
\citep{marigo, ekstrom}: ${T_{\mathrm{eff}}=30\,000\,\mathrm{K},M=50M_{\odot},R_{\star}=30R_{\odot}}$.
The calculations were extended up to a considerable distance from the parent star, although this may in most cases be 
a purely hypothetical issue due to the low disk density in such regions.
According to the analytical prescriptions introduced in Sects. \ref{baseq} and \ref{radthin}, we solve the set of
Eqs.~\eqref{massconserve}-\eqref{phimcon} with the use of Eq.~\eqref{angfl}.

The inner boundary (stellar equatorial) values were adopted 
on the equatorial radius of the critically rotating star ($R_{\text{eq}}=3/2\,R_{\star}$)
in following way. The estimation of inner boundary surface density $\Sigma(R_{\text{eq}})$ is 
implemented as a fixed boundary value; in the case of a critically rotating B0-type 
star the isothermal disk surface density is
$\Sigma(R_{\text{eq}})=1.6\times 10^2$ g cm$^{-2}$, roughly corresponding to
$\dot{M}\approx 10^{-9} M_{\odot}$ yr$^{-1}$ derived by
\citet{granada}. In the case of a critically rotating Pop III type star the isothermal disk surface density is
$\Sigma(R_{\text{eq}})=1.6\times 10^5$ g cm$^{-2}$, which roughly corresponds to
$\dot{M}\approx 10^{-6} M_{\odot}$ yr$^{-1}$ \citep{ekstrom}.
Similarly to the stationary calculations described in Sect.~\ref{statcalc}, the
time-dependent models are independent of the scaling of the surface density $\Sigma$.
The inner boundary condition for $V_R$ is free, i.~e. the quantity is extrapolated 
from mesh interior values as a 0-th order extrapolation.
As the inner boundary condition for $V_{\phi}$ we assume the Keplerian velocity (critically rotating 
stellar equatorial surface).
The outer boundary conditions are considered as outflowing for all quantities.

We set the initial surface density profile to $\Sigma\sim R^{-2}$ 
\citep{okazaki}. We start the numerical calculation with zero initial gas radial velocity and Keplerian rotational velocity
throughout the whole disk. 
Our numerical test confirmed that the final solution is independent of the
initial conditions and does not depend on the actual inner boundary condition
for $V_R$.

\section{Results of numerical models}

\subsection{Stationary calculations}
\label{statmod}

\begin{figure} [t]
\centering\resizebox{0.84\hsize}{!}{\includegraphics{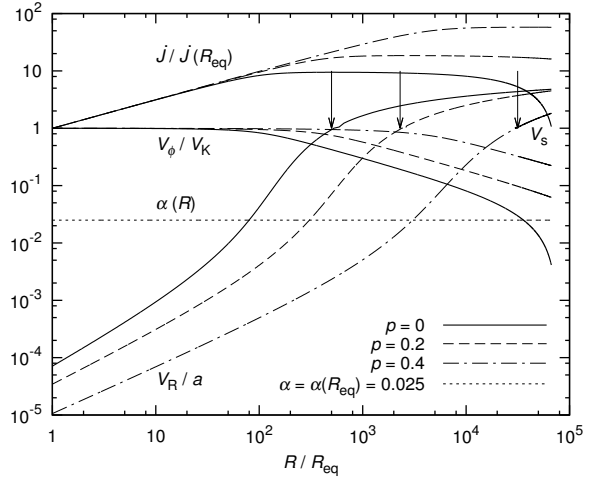}}
\caption{Dependence of the relative radial and azimuthal velocities and the relative angular momentum loss rate 
${\dot{J}/\dot{J}(R_{\text{eq}})}$ on radius
for various temperature profiles calculated by the method described in Sect.~\ref{statcalc}. 
Constant viscosity ${\alpha=0.025}$ is assumed. 
Arrows denote the sonic point.}
\label{fig1}
\end{figure}
\begin{figure} [t]
\centering\resizebox{0.84\hsize}{!}{\includegraphics{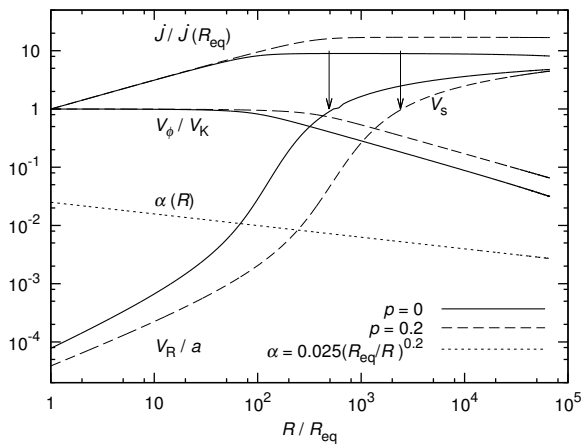}}
\caption{As in Fig.~\ref{fig1}, but
with variable ${\alpha}$ parameter, ${\alpha\sim R^{-0.2}}$.}
\label{fig2}
\end{figure}
In stationary calculations the first order linear viscosity was 
assumed according to Eq.~\eqref{viscon}. 
The model with constant viscosity profile and $p=0$
in Fig.~\ref{fig1}
shows a rapid decrease in the rotational velocity and the angular momentum loss at large radii. These quantities 
even drop to negative values in the case of the adopted 
first order linear viscosity prescription (see Sect.~\ref{radthin}). 
The velocity drop is caused by the increase of the second term in
Eq.~\eqref{angestim3} at large radii. We consider the drop to be unphysical.
As a solution to this problem within stationary calculations we introduce the models with
power law viscosity decline. Up to a certain value of $p$ parameter in temperature power law profile the models avoid the rapid 
rotational velocity drop (show 
constant angular momentum loss rate) in supersonic region (Fig.~\ref{fig2}). 

\subsection{Disk evolution time}
\label{chartime}

\begin{figure} [t]
\centering\resizebox{0.7\hsize}{!}{\includegraphics{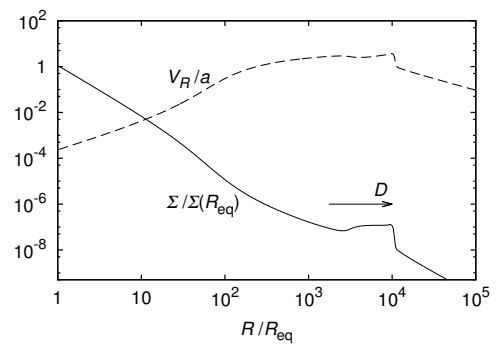}}
\caption{The snapshot of the radial velocity and the surface density transforming wave in the case of 
a B0-type star isothermal constant viscosity model (see Fig.~\ref{B0pt0}),
the time $t_{\text{dyn}}\approx 40\,\text{yr}$.
The wave propagation velocity is denoted as $D$.
In the model, the ratio $\Sigma_1/\Sigma_0$ (surface densities behind and in front of the wavefront) 
is about one order of magnitude and slightly increases with the distance.}
\label{shock}
\end{figure}

In the time-dependent 
models we recognize the wave that converges the initial state of calculated quantities to their final stationary state.
Because of the initial density distribution we may regard the wave as physical (not only numerical artefact).
We assume that during the disk developing phase a similar transforming wave occurs and
its amplitude and velocity depends on physical conditions (namely the density distribution) 
in the stellar surroundings. There might be a possibility to observe some 
bow shocks at the boundary between the developing disk and the interstellar medium  
(even though the disk radial velocity in the distant regions is about one order of magnitude 
lower than in the case of line-driven stellar winds).
The wave may also determine the timescale of the Be star disk growth and
dissipation phases \citep[e.g.,][]{guhaj,stebar}.
In the subsonic region the wave establishes nearly hydrostatic equilibrium in
the radial direction (Eq.~\eqref{radmomconserve}) and the wave speed approximately equals the sound speed.
In the supersonic region this wave propagates as a shock wave. Its propagation speed can be approximated
as $D=a\sqrt{\Sigma_1/\Sigma_0}$ (see Fig.~\ref{shock}), where the subscripts
$0$ and $1$ denote the values in front of and behind the shock front, respectively \citep{Zelda}. We regard 
the shock propagation time as the dynamical time $t_{\text{dyn}}\approx R/D=0.3R/a$. For example for the distance 
$10^4\,R_{\text{eq}}$ the isothermal constant viscosity B0-type star disk model gives $t_{\text{dyn}}\approx 40\,\text{yr}$.
The dynamical time is almost independent of the viscosity while it significantly increases with decreasing temperature.
\begin{figure}
\centering\resizebox{0.72\hsize}{!}{\includegraphics{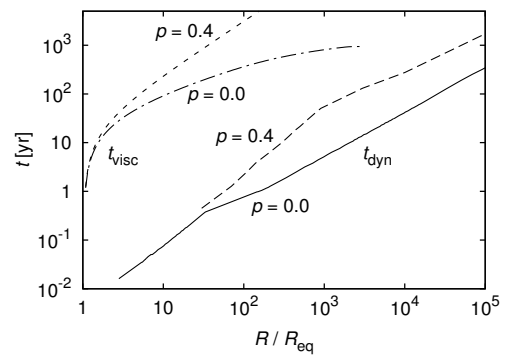}}
\caption{Comparison of the density wave propagation time (lower two branches denoted as $t_{\text{dyn}}$)
with the disk viscous time (upper two branches denoted as $t_{\text{visc}}$),
for B0-type star isothermal ($p=0$) constant viscosity model and the model with decreasing ($p=0.4$) temperature 
profile (see Fig.~\ref{B0pt0n04}), in dependence on radius.
The disk viscous time is calculated from Eq.~\eqref{visctime}. Since the rotational 
velocity $V_{\phi}$ is adopted from the models, the graph of $t_{\text{visc}}$ is cut off
in the region of the rapid rotational velocity drop. The plotted values of $t_{\text{dyn}}$ are adopted from the models.}
\label{shortlong}
\end{figure} 
We associate the disk evolution time with disk viscous time \citep{okazaki, maeder} 
\begin{equation}\label{visctime}
t_{\text{visc}}=\int_{R_{\text{eq}}}^RV_{\phi}\,\text{d}R/(\alpha a^2), 
\end{equation}
i.e., the timescale on which matter diffuses through the disk under the effect of the viscous torques \citep{Frank}.
In the isothermal constant viscosity case the same model gives 
$t_{\text{visc}}\sim 10^2\,\text{yr}$ for the sonic point radius.
The viscous time significantly grows with temperature and viscosity.
The comparison of the two times $t_{\text{dyn}}$ and $t_{\text{visc}}$ is in Fig.~\ref{shortlong}.

We also investigated the case with arbitrarily low nonzero initial surface density values in 
the whole computational domain.
We assumed the initial (Keplerian) value of $V_{\phi}$ up to only a few tens of stellar radii
followed by a discontinuous jump down to zero.
Even using these initial conditions the disk evolves to a large distance and it converges
to the proper final state. The density in the outer disk radius region
is lower than the initial density value, forming a rarefaction wave that really extends
radially with time (but more work on this point is needed). 

\subsection{Stationary state reached by time-dependent models}  
\label{timemod}

\begin{figure} [t]
\centering\resizebox{0.8\hsize}{!}{\includegraphics{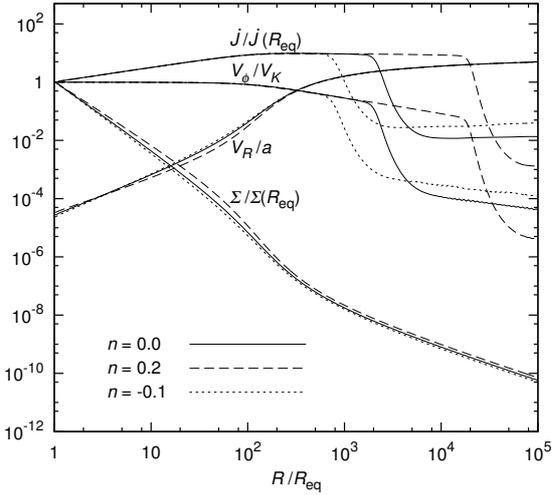}}
\caption{The dependence of scaled vertically integrated density and radial and azimuthal disk velocities and the scaled 
angular momentum loss rate  
$\dot{J}/\dot{J} (R_{\mathrm{eq}})$ on radius in the case of isothermal disk ($p=0$) of selected B0-type star
for various radial viscosity profiles (various $n$) in a final stationary state of
time-dependent models (the rapid drop of rotational velocity and the angular momentum loss rate in the outer disk region is 
a stationary jump, not a shock wave).}
\label{B0pt0}
\end{figure}

From the calculations it follows
that the profiles of surface density and radial velocity as well as the sonic point distance
(where $V_{R}/a=1$) very weakly depend on the viscosity parameter $n$.
The outer limit of Keplerian rotation velocity region ($V_{\phi}\sim R^{-0.5}$) is almost independent of the viscosity parameter. 
The calculations nevertheless show strong dependence of the outer edge of the region where the rotational velocity behaves as
angular momentum conserving ($V_{\phi}\sim R^{-1}$) (i.e., of the distance where the rotational velocity begins to 
rapidly drop) on viscosity parameter (for a given temperature profile).  
For a selected range of viscosity parameter $n$ the distance of this region differs approximately by
one order of magnitude (see, e.g., Fig.~\ref{B0pt0}). 
In the models the location of this rapid rotational velocity drop
does not exceed the radius where the disk equatorial density falls to 
averaged interstellar medium density (its mean density we assume as $10^{-23}\,\text{g}\,\text{cm}^{-3}$ \citep[e.g.,][]{misirio, maeder}).
At this distance a kinetic plasma modelling
would likely be required; moreover, the interaction of the disk with interstellar medium 
has to be taken into account.

Within time-dependent calculations we examined the differences in the
numerical results between the two different prescriptions for viscous torque (Eqs.~\eqref{viscon} and \eqref{angfl}).
Similarly to the steady disk calculations (see Sect.~\ref{statmod}), the
first order linear viscosity calculations (Eq.~\eqref{viscon}) exhibit the rapid rotational velocity 
drop to negative values and consequently indicate very slow convergence to zero. 
The rotational velocity profiles calculated involving
the second order linear viscosity term (Eq.~\eqref{angfl}) confirm the analytical 
result $V_{\phi}>0$ throughout the entire disk range (see Sect.~\ref{radthin}).

Figure \ref{B0pt0} illustrates the isothermal case ($p=0$) of a B0-type star 
($\dot{M}\approx 10^{-9}\,M_{\odot}$ yr$^{-1}$, see Sect. \ref{tdepcalc})
with $\alpha(R_{\text{eq}})=0.025$ \citep{Penna}.
The calculated radius of the sonic point $R_\text{s}\approx 550\,R_{\mathrm{eq}}$ roughly corresponds to
the analytical prediction from Eq.~\eqref{sonicestR} with
$R_\text{s}$ being approximately $480\,R_{\mathrm{eq}}$.
The maximum angular momentum loss rate $\dot{J}_{\text{max}}$ (see Eq.~\eqref{sonicestL}) is independent of viscosity 
while it strongly depends on the profile of temperature.
Since $\dot{J}_{\text{max}}$ roughly equals the angular momentum loss rate at the sonic point 
radius, we assume the total angular momentum contained in the disk to be
\begin{equation}\label{angtot}
J_{\text{disk}}=\int_{R_{\text{eq}}}^{R_{\text{s}}}2\pi R^2\Sigma V_{\phi}\,\text{d}R,
\end{equation} 
and we therefore regard $R_\text{s}$ as the disk outer edge. Analogously it also determines the mass of the disk
\begin{equation}\label{masstot}
M_{\text{disk}}=\int_{R_{\text{eq}}}^{R_{\text{s}}}2\pi R\Sigma\,\text{d}R.
\end{equation}
Comparing for example
the total disk angular momentum $J_{\text{disk}}$ with the total stellar angular momentum $J_{\star}=\eta 
MR_{\star}^2\Omega_\text{crit}$ where 
a nondimensional parameter $\eta=0.05$ \citep{Meynet1}, in this case the ratio $J_{\text{disk}}/J_{\star}=1.2\times 10^{-6}$.

\begin{figure} [t]
\centering\resizebox{0.8\hsize}{!}{\includegraphics{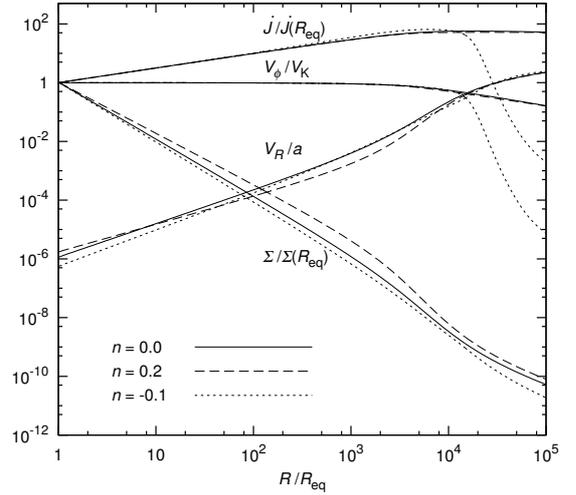}}
\caption{As in Fig.~\ref{B0pt0}, however for temperature decreasing with a power
law with $p=0.4$.
Inner boundary viscosity $\alpha(R_{\text{eq}})=0.025$ is considered. 
The characteristic radii (sonic point distance, 
outer disk radius) are in this case significantly larger.} 
\label{B0pt0n04}
\end{figure}

Figure \ref{B0pt0n04} shows the case of decreasing temperature profile ($p=0.4$) of the 
B0-type star
with the same viscosity profiles as in Fig.~\ref{B0pt0}.
The sonic point distance is roughly
$R_{\text{s}}\approx 31\,500\,R_{\mathrm{eq}}$ for all the viscosity profiles, which is about two orders of magnitude larger
than in the isothermal case. 
In this model the ratio $J_{\text{disk}}/J_{\star}=7.9\times 10^{-5}$,
which is similarly 
about two orders of magnitude larger than in the isothermal case.

\begin{figure} [t]
\centering\resizebox{0.8\hsize}{!}{\includegraphics{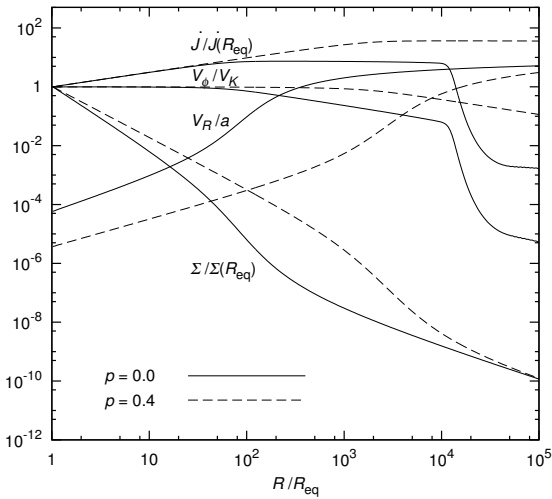}}
\caption{Comparison of the radial profiles of relative surface density and relative velocities and the radial
profiles of angular momentum loss in the case of decreasing viscosity ($n=0.2$) for isothermal disk ($p=0$)
and for outward decreasing temperature profile ($p=0.4$) in a final stationary state of time-dependent disk models
for Pop III star.}
\label{ptI}
\end{figure}

Figure \ref{ptI} shows the calculated profiles of the pop III star's disk 
($\dot{M}\approx 10^{-6}\,M_{\odot}$ yr$^{-1}$, see Sect. \ref{tdepcalc}) with various temperature parameters 
for fixed decreasing viscosity ($n=0.2$).
The graph clearly shows a strong dependence of radii of the sonic point and of 
the rapid rotational velocity drop, as well as of the slopes 
of surface density and radial velocity on temperature profile.
The sonic point radius is located at
$R_{\text{s}}\approx 360\,R_{\mathrm{eq}}$ in the isothermal case and
$\approx 16\,500\,R_{\mathrm{eq}}$ in case of radially decreasing temperature with selected parameter $p=0.4$. 
The use of the same dimensionless parameter $\eta=0.05$ 
gives the ratio $J_{\text{disk}}/J_{\star}=1.2\times 10^{-3}$ for the isothermal model and
$J_{\text{disk}}/J_{\star}=0.2$ for the model with decreasing temperature.
In the latter case the disk carries away a significant fraction of stellar angular momentum
and the star may not have enough angular momentum to develop the disk fully.
In this case the stellar evolution has to be calculated together with the disk evolution.
The calculations support the conclusion that the unphysical drop of the
rotational velocity can be avoided in the models with radially decreasing
viscosity parameter and temperature.

Within the time-dependent calculations we also examined 
subcritically rotating stars modifying the boundary condition for $V_{\phi}$.
For the inner boundary value of the azimuthal velocity
$V_{\phi}(R_{\text{eq}})\gtrapprox 0.97 V_K(R_{\text{eq}})$
(where $V_K(R_{\text{eq}})$ denotes the Keplerian velocity at the stellar equator),
the models precisely converge in the supersonic region. However, in the case
when the boundary rotational velocity is only slightly higher than the above limit 
($0.97 V_K(R_{\text{eq}})\lessapprox V_{\phi}(R_{\text{eq}})\lessapprox 0.98 V_K(R_{\text{eq}})$) 
there occur (more or less regular) pulsations in the density and radial velocity profiles in the region close to the star.
For $V_{\phi}(R_{\text{eq}})\lessapprox 0.97 V_K(R_{\text{eq}})$ the density (and consequently the radial velocity) 
profile is unstable and gradually declines; for lower $V_{\phi}(R_{\text{eq}})$ the decrease in density is faster. 

\section{Conclusions}
\label{conc}

We calculated axisymmetrical, vertically integrated one-dimensional time-dependent models of decretion disks of
critically rotating stars.
For this purpose we developed a numerical code for time-dependent
hydrodynamical modelling that includes full Navier-Stokes viscosity.
We extrapolate the disk temperature profiles obtained by NLTE simulations
\citep[e.g.][]{Carcio}
by the parameterized profiles. Various temperature profiles give different slopes of integrated density decrease
throughout most of the disk.
The radial dependence of the disk equatorial density may in various regions differ
from the parameterized density profiles generally used in models dealing with the disk thermal structure \citep[e.g.,][]{Sigut2009, McGill}.
Since the radial profile of the $\alpha$ viscosity parameter
is not quite certain, we parameterize it via an independent power law radial dependence.

The time-dependent one-dimensional models 
confirm the basic results obtained in the stationary models 
in respect of the sonic point distance and of the distance of the disk outer edge (i.e., the radius where the
rotational velocity begins to rapidly decrease) on parameterized temperature and viscosity profiles. 
The sonic point is located at larger radii in the models with steeper temperature decrease 
while its radius very weakly depends on the viscosity profile. The sonic radius
strongly depends on both temperature and viscosity profiles
and does not exceed the distance where we expect the disk equatorial density may drop to the average density of the interstellar medium.
Consequently, the total angular momentum 
contained in the disk and the mass of the disk increase with the decreasing temperature and viscosity profiles.
The analytical relations provided in Sect. \ref{radthin} give adequate approximations of the numerical models.
The unphysical drop of the rotational velocity and angular momentum loss at
large radii, which is present in the isothermal models with constant viscosity
parameters, can be avoided in the models with decreasing temperature and
viscosity parameters.

\begin{acknowledgements}
The access to computing and storage facilities owned by parties and projects contributing to the 
National Grid Infrastructure MetaCentrum, provided under the program 
"Projects of Large Infrastructure for Research, Development, and Innovations" (LM2010005) is appreciated.
This work was supported by the grant GA \v{C}R 13-10589S.
\end{acknowledgements}

\end{document}